\documentclass[12pt]{article}
\usepackage[utf8]{inputenc}
\usepackage[breaklinks]{hyperref}
\usepackage{microtype}
\usepackage{graphicx}
\usepackage{amsmath}

\def\d{\delta}

\def\s{\sigma}
\def\b{\beta}

\begin{document}
\begin{center}
{\bf Evolution of cooperation in spatially structured group-selection models of the continuous Prisoner's Dilemma} \\

\bigskip
Yaroslav Ispolatov\footnote{Department of Physics, Center for Interdisciplinary Research in Astrophysics and Space Science, University of Santiago of Chile, jaros007@gmail.com},
Burton Simon\footnote{Department of Mathematical and Statistical Sciences, University of Colorado Denver}, and Michael Doebeli\footnote{Department of Zoology and Department of Mathematics, University of British Columbia}
\end{center}

\bigskip
\begin{center}
{\bf Abstract}
\end{center}
{\small
We consider continuous Prisoner's Dilemma games played in a spatial setting by
group-structured populations. The evolutionary dynamics is driven by within-group individual-level birth and death events, and by group-level fission and
extinction events. Within groups, individuals play well-mixed games, while
groups play games on a 1-dimensional spatial grid against their nearest neighbours.
Payoffs from individual-level
games affect birth rates of individuals, and payoffs from group-level
games affect group extinction and fission probabilities.
It is well known that the within-group
evolution by itself always results in a complete loss of
cooperation, and that in the group-level games, defection is also favoured if these games are played in well-mixed populations of groups.
Here we show that despite this double disadvantage, cooperation can be maintained due to spatially structured group-level dynamics.
Mirroring results from one-level selection models, the spatial nature of games between groups and the resulting fissioning and extinction events is essential for the evolution of cooperation in these models, but needs to manifest itself in specific ways in order to be effective: we find that
higher levels of cooperation evolve when the selection generated by games between groups acts
locally rather than globally.    
   
}


\section {Introduction}
\begin{sloppypar}
Evolution of  cooperation has been an important
topic in biology and social sciences \cite{axelrod1981evolution}. Game theoretical methods are well suited and have been widely used 
to investigate this problem \cite{hofbauersigmund2012games, sigmund1999evolutionary, traulsen2023future}. The simplest  and most iconic example of such games is the
Prisoner’s Dilemma game. A common version of this game has individuals contribute to a common good while incurring costs for such contributions \cite{doebeli2005models}. With well-mixed interactions (mean field case), this makes the evolutionary loss  of cooperation inevitable, which allows for investigations of various modifications that can explain the maintenance of cooperation in this game. 

A notable example of such an extension is the spatial structuring of populations, where games are played locally among close neighbours \cite{nowak1992evolutionary, nowaksigmund2000grids}. It has been shown that both in the classic discrete Prisoner’s Dilemma with two strategies, full cooperators and full defection, and in the continuous Prisoner’s Dilemma game, in which the level of cooperation can vary continuously between 0 and an upper limit, cooperation can evolve due to formation of spatially segregated domains \cite{nowak1992evolutionary, nowaksigmund2000grids,doebeli2005models}. In these domains, the contributions of cooperators are shared predominantly among fellow cooperators, resulting in higher birth or lower death rates for cooperators. 

\end{sloppypar}
Another extension that allows for the emergence of cooperation is the introduction of multiple levels of selection \cite{kerrgodfrey2002mls}. In such models, individuals playing the Prisoner’s Dilemma game are assumed to live in groups, within which they play the game and obtain game payoffs. If one then assumes that groups with higher average within-group payoffs ``fare better'' in some sense that has to be specified, then cooperation can be maintained, because even though cooperation tends to be lost within each group, groups with higher average levels of cooperation do better than groups with lower average levels of cooperation. Again, the maintenance of cooperation in such multi-level selection models depends on the details implemented in these models. Most often, the between-group selection has been assumed to depend on simpler mechanisms than the within-group selection, e.g. by assuming that selection at the group level depends on aggregate characteristics, such as the above-mentioned within-group level of cooperation, or the group size {  \cite{akdeniz2020cancellation, cooney2022long, cooney2019replicator, traulsen2006evolution, luo2014unifying, cooney2023evolutionary, fog2026simulation}}.  In contrast, we have recently studied several multi-level models in which games are played not only within groups, but also between groups, so that between-group games determine group-level events such as fissioning and extinction \cite{simon2024fission,  simon2024evolutionary}. In the case where the game played both within-groups and between groups is the Prisoner’s Dilemma game, and with well-mixed (mean field) interactions at both the individual and the group level, it is easy to see that cooperation cannot evolve, because it is selected against at both the individual and the group level.

In the present paper, we extend the multi-level selection models of \cite{simon2024fission,  simon2024evolutionary}  to investigate the effect of spatial structure on the evolution of cooperation in the multi-level Prisoner’s Dilemma game. Spatial structure is implemented by assuming that groups of individuals occupy a 1-dimensional spatial grid and that group-level events are local. In particular, group-level games are played between neighbouring groups. The model will be defined mathematically in the next section. Apart from its spatial structure and update rules, it is similar to the models used to study group selection in \cite{simon2010dynamical,simon2013towards,simon2016group, simon2024fission, simon2024evolutionary}:
\begin{itemize}
    \item Individuals play well-mixed games against other members of their group; groups play games against their neighbouring groups. 
    \item Individual birth rates depend on payoffs from the within-group games (higher payoffs lead to higher birth rates). Individual death rates depend on the size of the group.
    \item Strategies in the between-group games are the average level of cooperation of the individuals within each group.  
    \item Groups are updated according to  payoffs of between-group games. The group chosen for reproduction randomly fissions into two daughter groups, replacing itself and the group targeted for extinction. 
\end{itemize}

It is known from models of single-level spatial games (i.e., spatial games without group selection) that the details of the implementation of update events in the spatial grid are important and can be crucial for the maintenance of cooperation.   
Specifically, in single-level spatial games there are birth and death events, and the ordering of these events in conjunction with the assumptions about which events are affected by game payoffs is crucial for the dynamics of the spatial game. For example, if payoffs affect birth rates, whether death of an individual is followed by birth of another individual (DB) or vice versa (BD) determines whether cooperation can evolve or not in the spatial (discrete) Prisoner’s Dilemma. In its simplest form, the DB scenario is implemented as the replacement of a random player by a copy of its best-performing neighbour, and it is well known that cooperation can persist in this case \cite{ohtsuki2006simple, taylor2007evolution, zukewich2013consolidating, hauert2021spatial}. However, in the BD scenario, which is usually implemented as the replacement of a random neighbour  of the globally best-performing player by a copy of that best performer, cooperation does not evolve \cite{ohtsuki2006simple, taylor2007evolution, zukewich2013consolidating, hauert2021spatial}. The reason for this, also known as the ``cancellation effect'' {  \cite{akdeniz2020cancellation}}, is the inability of cooperators to ``export'' their cooperation to other regions of the system under BD update rules \cite{ohtsuki2006simple, taylor2007evolution}. Since the best performing players are those within the domains of cooperators, their neighbours are often cooperators as well, so the BD updates tend to replace cooperators by cooperators. At the same time, these domains and the interfaces between the domains of high and low cooperation (defection) are susceptible to the invasion of defectors \cite{ohtsuki2006simple}. 

We find that similar mechanisms are in play in our spatially structured multi-level Prisoner’s Dilemma models. We show that under a suitable balance between the strength of within-group and between-group selection,  spatial structure can indeed rescue cooperation against the double disadvantage of cooperation in both the within-group and the between-group Prisoner’s Dilemma game. But we also find that the details of the update rules are important: update rules that are local in the sense that they depend on the payoffs of  neighbours of a randomly chosen group are more important for maintaining cooperation than global update rules, in which the system-wide best or worst group is chosen for birth or death. In short, while local selection is essential for the maintenance of cooperation, global selection is not. Thus, our results extend well-known results from single-level spatial games to multi-level selection models in which games are played both between individuals and between groups.

\section {The Model}
We study a spatially structured population of groups, with each group consisting of individuals that are characterized by their level of cooperation $x\in [0,1]$. Within each group, individuals play well-mixed continuous Prisoner’s Dilemma games, in which the payoff to an individual with strategy $x$ playing an individual with strategy $y$ is 
\begin{align}\label{psi}
P(x,y) = By-Cx,
\end{align}
where the parameters $B>0$ and $C>0$ reflect the benefit and the cost of cooperation, respectively. Generally speaking, with random interactions cooperation is doomed in this game, because the only way players can affect their payoff is by lowering costs, i.e. by lowering their cooperation level. We note that for this game to be a ''social dilemma'', in which payoffs in a (monomorphic) population of cooperators are higher than in a population of defectors, requires $B>C$. 

In our set-up, an individual’s birth rate depends on its expected payoff from games with all other individuals in its group. 
An individual's death rate is proportional to the size of its group, constraining the group size. The resulting within-group birth-death process determines the within-group dynamics in the model.

There is also a birth-death process at the group level, at which birth is represented by fissioning and death by group extinction. Groups are arranged on a 1-dimensional grid with periodic boundaries (a ring) and receive payoffs from games played against their nearest neighbours to the left and to the right. In these games, each group's strategy $\bar x\in [0,1]$ is the average of the strategies of all individuals in the group.
Although an individual's level of cooperation does not change during its lifetime, 
group traits change in time due to internal individual-level  birth and death  events. Payoffs from games between groups affect the group-level events, fissioning and extinction (details below).
When a group fissions, it breaks randomly into two pieces, which replace the ancestral group and its neighbour marked for extinction.

\subsection{Within-group dynamics}
The exact simulation of all within-group and between-group events is slow due to the large number of events.
 We therefore use the ``hybrid'' method for simulating group-structured populations described in \cite{simon2012numerical, simon2024fission, simon2024evolutionary}.
In the hybrid method the state of a group is approximated by a continuous within-group population density distribution, whose deterministic dynamics are described by a partial differential equation (PDE), while all group-level events are described explicitly, with groups participating as discrete entities. The PDE  governing within-group population dynamics is derived from the birth and death rate functions and is justified by large population limits  \cite{champagnat2006unifying}. 
In our numerical experiments the groups are fairly large (around 100 individuals on average) so the hybrid method is appropriate. 
Using the PDE,  the state of a group becomes a continuous density, $n(x),~x\in[0,1]$, {  and the size of the group $\eta$ is }
\begin{align}\label{eta}
\eta = \int_0^1 n(x)dx 
\end{align}
In the continuous approximation of the within-group population dynamics, the per capita birth rate $b(x)$ becomes
\begin{equation}\label{birthrate2}
b(x) = \beta \frac{e^{s\psi(x)}}{\eta^{-1}\int_0^1 e^{s\psi(y)}n(y) dy},
\end{equation}
where
\begin{equation}\label{psi2}
\psi(x) =\eta ^{-1}\int_0^1 P(x,u)n(u) du
\end{equation}
is the expected payoff to an individual with type $x$ from  games against all  individuals $u$ in the group, where each game yields a payoff $P(x,u)$ given by eq. (1) for the continuous Prisoner’s Dilemma.
The payoff is normalized by the size of the group $\eta$.
{  The parameter $\beta$ in (\ref{birthrate2}) defines the scale of the birth rate} and  $s\geq 0$ defines the strength of selection, so that within-group evolution is neutral for $s=0$. 
 (Note that the strength of selection is defined by the combination of $\beta$ and $s$, so the small values of $s$ used in our simulations should not be interpreted as very weak selection, as they are compensated by the large values of $\beta$, see the Appendix).  
The game payoffs are exponentiated (so that birth rates are positive for any game payoffs) and normalized to ensure that the average per capita birth rate in each group is always equal to $\beta$, and in particular is independent of the game played. This allows us to separate the dynamics of the total within-group population size $\eta$ from the evolutionary dynamics of the cooperation level $x$.

The per capita death rate in a group with $\eta$ individuals does not depend on $x$ and is proportional to the group size,
\begin{equation}\label{deathrate}
D = \delta \eta.
\end{equation}

These birth and death rates define the PDE for the within-group population density $n(x,t)$:
\begin{equation}\label{integdiffeqn}
\frac{\partial n(x,t)}{\partial t} = \left[b(x) - D\right]n(x,t) + R\frac{\partial^2 n(x,t)}{\partial x^2} 
\end{equation}
where a diffusion term with a small coefficient $R$ is added to mimic mutations in $x$ \cite{champagnat2006unifying}. We note that with this setup, the total population size of each group equilibrates at the value $\eta=\beta/\delta$. Except for mutations, the cooperation level $x$ converges to 0 (see the Appendix), as the payoff (\ref{psi}) of the continuous Prisoner’s Dilemma dictates that lower cooperation levels always lead to higher expected payoffs, and hence higher per capita birth rates (\ref{birthrate2}). The speed of this convergence is gauged by the parameter $s\beta$.

To implement the PDE numerically, we approximate the individual-type space, $[0,1]$, by a set of $m$ equally sized bins between 0 and 1. For each group, this leads to a system of $m$ nonlinear differential equations for the populations of each bin, which are  solved by standard numerical techniques \cite{ simon2024fission, simon2024evolutionary}. 

\subsection{Between-group games and the group-level update rules}
The group-level events are discrete with the dynamics determined by the payoffs of between-group games, which in their turn depend on the within-group levels of cooperation $\{\bar x \}_{i}$. {  (We denote group-level variables by curly brackets with the subscript denoting the group).} Groups are positioned on a circle and play games with their nearest neighbours on both sides, resulting in a payoff to group $i$
\begin{align}
\label{group_type}
P_i=B (\{\bar x \}_{i-1}+\{\bar x \}_{i+1})/2 - C \{\bar x \}_{i}. 
\end{align}
For simplicity, we assume that the groups play the same continuous Prisoner’s Dilemma game as individuals within groups. In particular, we note that if this between-group game is played in well-mixed populations of groups, cooperation is doomed even in the absence of within-group selection, because the payoff structure dictates that lower $\bar x$ always yield higher payoffs, regardless of the group traits of other groups.

Groups fission and go extinct and since the number of groups is set to be constant, each fissioning event is accompanied by an extinction event (or vice versa). A target group fissions randomly with one daughter group staying at the same site while the other daughter group replaces one of its neighbours. The fissioning scenario is similar to that implemented in our previous work, \cite{simon2024fission}, were we explored the effect of fissioning randomness on the group-level evolution.  To implement fissioning, we use the same binning of the within-group distribution of $x$ as we use for the numerical implementation of the within-group PDE. If $w_k(t), k=1,\ldots,m,$ denote the population size in the $m$ bins of the target group at the time of fissioning, each $w_k(t)$ is split randomly between the corresponding bins of two daughter groups.  
Let $\hat w_k\in [0,w_k]$ be the number of individuals in the $k$th bin of the parent group that go to the $k$th bin of daughter group 1, while the remainder go to daughter group 2. We choose 
\begin{align}
\label{fiss}
  \hat w_k= \hat {\mathcal {N}}\left( \frac{w_k}{2}, \frac{r^2 w_k}{4}\right ),
\end{align}
where $\hat {\mathcal {N}} (\mu,\sigma^2)$ is a normally distributed random variable, confined between $0$ and $2\mu$, and $r$ is the fission randomization scale.  We showed in \cite{simon2024fission} that for small $r$, when the normal distribution of $ \hat w_i$ is unperturbed by the cutoff $0 \leq \hat w_k \leq w_k$, such fissioning results in the variance between two daughter groups immediately after fissioning being proportional to $r^2$. $r=1$ is a continuous analog of individuals flipping a coin to choose a daughter group to join. The only difference from the setup used in \cite{simon2024fission} is that here we consider the same $m$ bins both for the numerical solution of (\ref{integdiffeqn}) and for fissioning.

\begin{figure}
   \includegraphics[width=.95\linewidth]{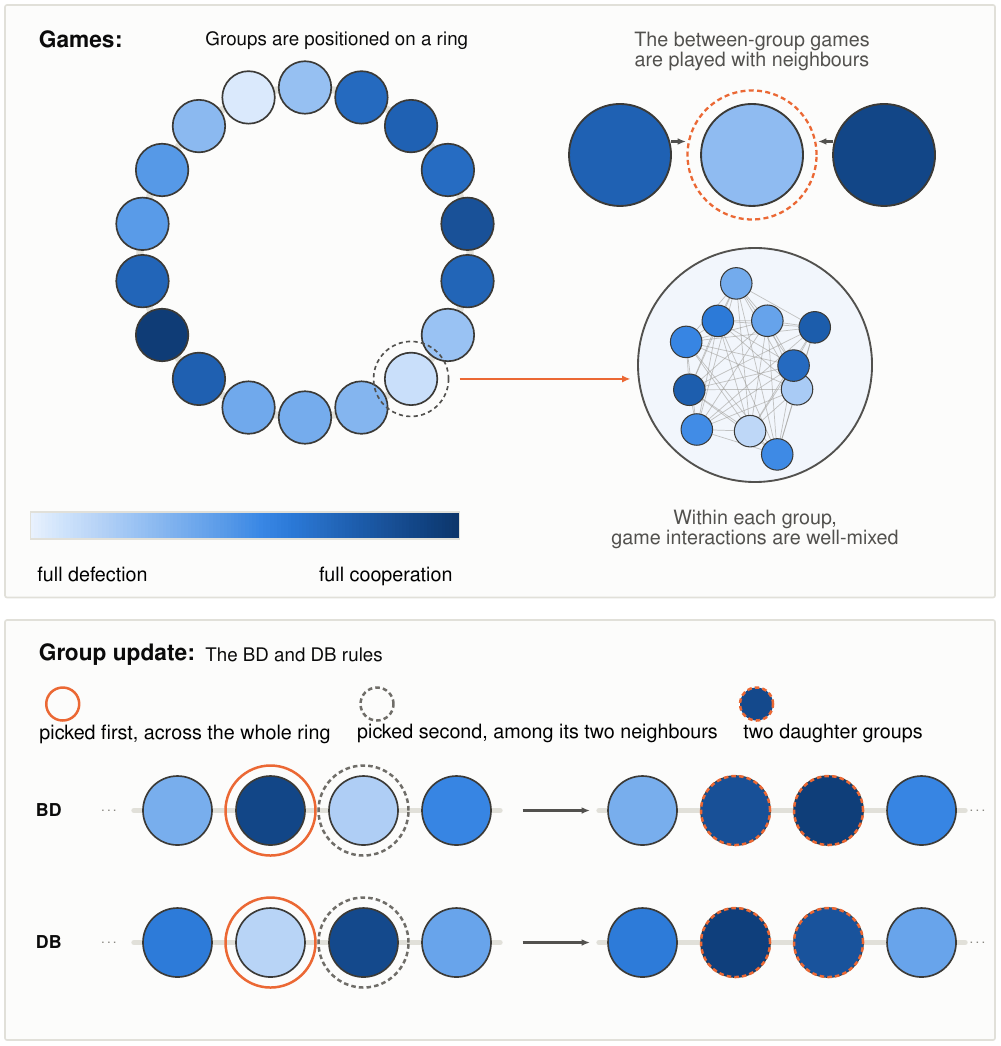}  
   
  \caption{  Schematic representation of the model: The top row  illustrates between- and within-group games; BD and DB update rules are sketched in the bottom row.} 
   \label{f0}
\end{figure}

{  The key to the group-level events are the update rules. Generally, these consist of two steps. First, a global target group is chosen (either for fissioning or extinction), then a local neighbour of the target group is chosen: if the global target group fissions, the local group is chosen for extinction and replaced by a daughter group of the fissioning global target group; if the global target group goes extinct, the local target group is chosen for fissioning, with one of the daughter groups replacing the extinct global targert group.  For single-level games two update schemes are traditionally considered, the ``Birth-Death''  rule and the  ``Death-Birth''  rule. In the terminology of \cite{traulsen2023future}, these rules are denoted by Bd and dB, respectively. In Bd the global target (a group in our case) has the highest payoff  in the whole population and is chosen to reproduce and have a daughter replace a randomly chosen neighbour. In dB the global target is chosen randomly and is replaced by a daughter of the neighbour with the highest payoff. Thus, capitalization (or not) in the notation refers to whether selection operates (or not) in the corresponding update step. Here we use generalized versions of these traditional update schemes, in which selection can act in both update steps (and in which the tradiatonal Bd and dB updates are special cases). In accordance with \cite{traulsen2023future}, these rules are denoted by BD and DB, respectively. We consider two parametrized families of scenarios:
\begin{enumerate}
\item {  Birth-Death (BD)} update. The group with the globally higher payoff fissions into the spot of its worst performing neighbour. 
\item {  Death-Birth (DB)}  update. The group with the globally lower payoff is goes extinct and is replaced by its highest-payoff neighbour. 
\end{enumerate}

  For global selection, i.e., for the choice of the global target group, we use a  parametrization of the strength of selection, ranging from weak to strong. Specifically, global selection is parametrized by a scaling constant $\s$ in the relation between the probability to be chosen as the target group, $Prob_i$, and the group's payoff, $P_i$. In the BD scenario, in which  better-performing groups are chosen, we have
\begin{align}
\label{softmaxb}
Prob_i=\frac{e^{\s P_i}}{\sum_je^{\s P_j}},
\end{align}
and in the DB scenario, in which worse-performing groups are chosen, we have
\begin{align}
\label{softmaxw}
Prob_i=\frac{e^{-\s P_i}}{\sum_je^{-\s P_j}}.
\end{align}  
This probabilistic targeting is often called the "Softmax" function, where by similarity with the Boltzmann distribution,  $\s$ is referred to as the "inverse temperature".
When $\s$ is small, global selection is close to neutral, i.e., selection of the target group is almost random, with $\s=0$ corresponding to neutral global selection. Thus, in BD, $s=0$ corresponds to the traditional bD, and in DB, $s=0$ corresponds to the traditional dB.
When $\s$ is large, that is, $\s P\gg1$, the absolutely best or worst performing group is chosen in the global update step. In simulations where the number of groups is inevitably finite, the $\s \rightarrow \infty$ limit often results in artifacts such as the persistent targeting of the same best or worst group. The constant $\s$ should be thought of as the strength of global group-level selection.  

 We did not parametrize the local step in the above rules, i.e., for local selection, which occurs among two neighbours, we simply select the worst-performing neighbour of the target group in BD updating and the best-performing neighbour of the target group in DB updating  (with random tie-breaking). The only exception to this occurs in cases when we consider neutral local selection, i.e., a random choice of the corresponding neighbour of the target group. Again following the notation in  \cite{traulsen2023future}, we denote cases with random local choice with Bd and Db, respectively, with the lower case letters standing for neutral selection. Fig. \ref{f0} provides a schematic depiction of the model setup, illustrating local games at the group level, well-mixed games within groups, and the group-level update rules. }

\subsection{Simulations}
The numerical simulations of the entire dynamics work as follows.   $G$ groups are arranged on a circle and their number stays constant throughout the simulation. In each group, bin populations $w_j$ are seeded randomly with the same average, so the initial states of all groups are statistically identical and the average initial cooperation level is $\langle \bar x \rangle \approx 1/2$.

The simulation proceeds with the numerical integration of a discretized version of Eq. (\ref{integdiffeqn}) with a fairly small timestep. The scales $\beta$ and $\delta$ of the within-group population dynamics are chosen to be large, so that the total within-group population sizes relax to  their equilibrium size after initiation or fissioning very fast compared to the {  within-group} evolutionary timescale (which is determined by both $\beta$ and $s$, or their product $\beta s$ in the limit of weak selection). Once every $1/G$ time units, all groups play games with their neighbours and their payoffs are computed. This time increment means that on average each group is updated once per time unit. {  This makes it possible to compare the strengths of within-group and between-group selection and also insures that the within-group population sizes reach their value before the next fissioning event.}  Depending on the update rules, the target group is selected with the probability given by (\ref{softmaxb}) or (\ref{softmaxw}), and the neighbour with the lowest or highest payoff is chosen, followed by extinction and fissioning according to (\ref{fiss}). The simulations run up to time $T$. The principal values of the various parameters we used are summarized in Table 1.  Any changes of parameters, necessary to illustrate the main points of the work, are described in the text.  
\begin{table}[h!]
  \begin{center}
    \caption{Baseline values of parameters used in simulations.}
    \label{table1}
    \begin{tabular}{|l|c|c|c|} 
            \hline  
      \textbf{Parameter} &  \textbf{Meaning} & \textbf{Value}\\
      \hline
$\b$ & Birth rate coefficient & $10^3$ \\
     \hline
$\d$ & Death rate coefficient & $10$ \\
     \hline
$s$ & Selection strength coefficient & $10^{-4}$ -- $10^{-3}$\\
      \hline  
$B$ & Game benefit coefficient & $2.5$\\
      \hline  
$C$ & Game cost coefficient & $1$\\
      \hline  
 $R$ & Mutation coefficient & $10^{-6}$\\
      \hline  
 $r$ & Fission randomization coefficient & $0.5$ -- 2\\
      \hline  
$\s$ & Group selection coefficient & 0 -- 4\\
      \hline  
 $m$ & Number of discretization bins in groups & $10$\\
      \hline  
 $G$ & Number of  groups & $10^{3}$\\
      \hline  
$T$ & Simulation time & $5\times10^{4}$\\
      \hline  
    \end{tabular}
  \end{center}
\end{table}

\section{Results}
\subsection{Bd update}
\begin{figure}
\centering
    \includegraphics[width=.95\linewidth]{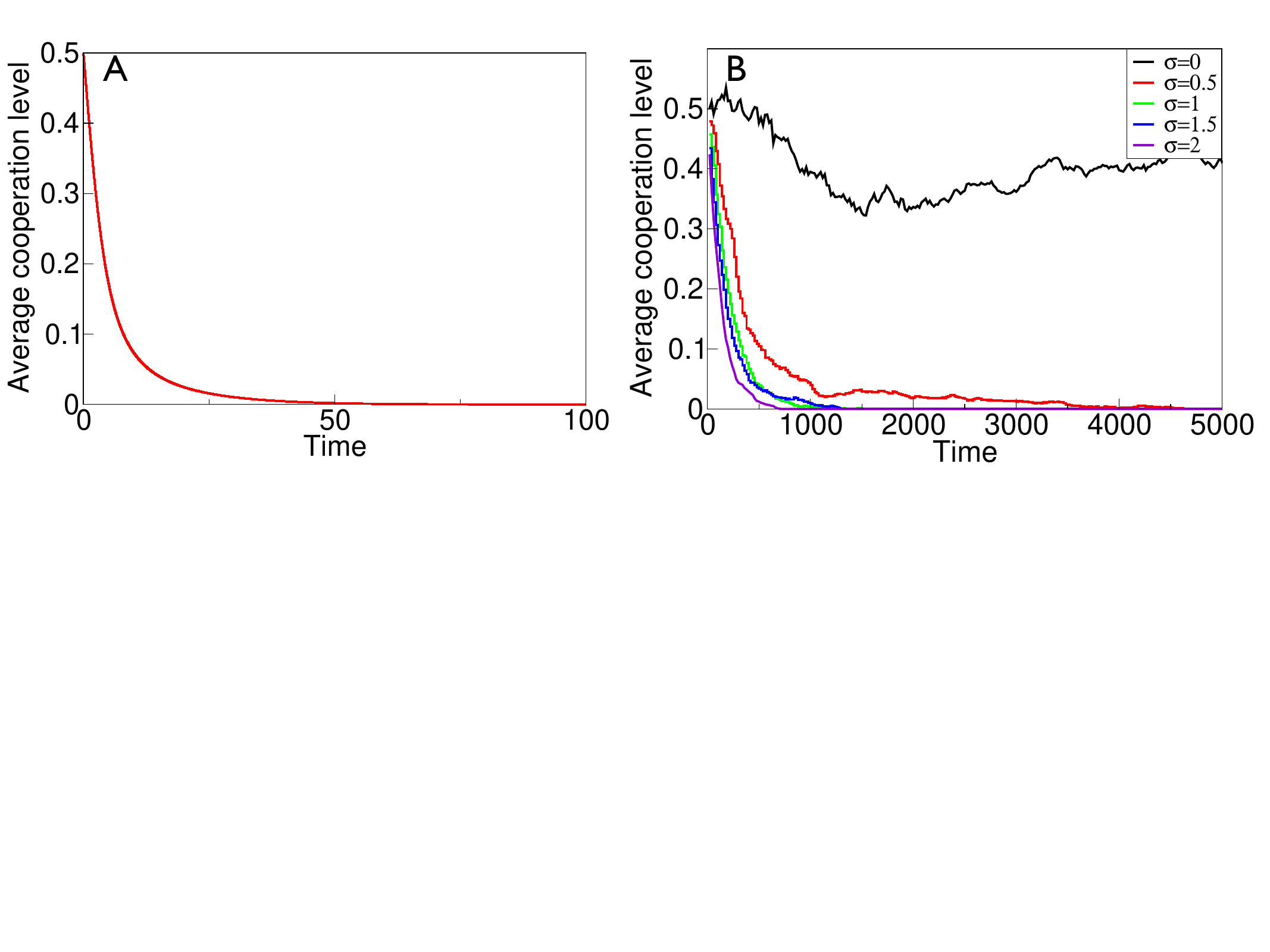} \\
   \caption{ The figure shows that cooperation cannot be maintained in two baseline scenarios. Left panel: In the absence of between-group games, the within-group games result in  a rapid decay of the cooperation level. Here $s=10^{-3}$. Right panel: the classic Bd between-group  update does not result in the evolution of cooperation even in the absence of within-group selection ($s=0$). The different curves shown correspond to specific values of the group-level selection coefficient $\s$. Note that in the case of  no selection ($\s=0$, black curve) the choice of target group is equiprobable, so the game payoff is irrelevant for the update and the average cooperation level  fluctuates around its initial value $\bar x \approx 0.5$ due to randomization during fissioning (here $r=1$).
  } 
   \label{f1}
\end{figure}

We first note that in the absence of between-group updates, cooperation can never establish (Fig. \ref{f1}, left panel), because the within-group Prisoner's Dilemma is played in well-mixed populations and hence selection favours ever lower cooperation levels. In the Appendix we show that the rate at which the average within-group cooperation level $\bar x$ (the first moment of Eq. (\ref{integdiffeqn}))  decays to zero is proportional to $\beta s$. A similar  derivation confirms that the between-group Prisoner's Dilemma played without any spatial structure, that is, in well-mixed populations of groups, is equally unable to maintain cooperation. 

Turning on the between-group games, we first investigate whether the group-level Bd update, i.e., the BD update with neutral local selection in the second update step, results in the maintenance of cooperation. We find that even in the absence of within-group selection ($s=0$), cooperation cannot be maintained in this case, Fig. \ref{f1}, right panel. This is expected because without the within-group games our model is essentially a one-level process, for which this result is well known \cite{ohtsuki2006simple, taylor2007evolution, zukewich2013consolidating, hauert2021spatial}. There is still some novelty in this result as the group fissioning process is not identical to the mutations in single-level games and in principle could have maintained cooperation by increasing the between-group variance. Note also that a more strict targeting of the best group for fissioning (larger $\s$) results in a faster decrease of cooperation compared to a more random (smaller $\s$) targeting.

\subsection {General BD update}
\begin{figure}
\centering
  \includegraphics[width=.95\linewidth]{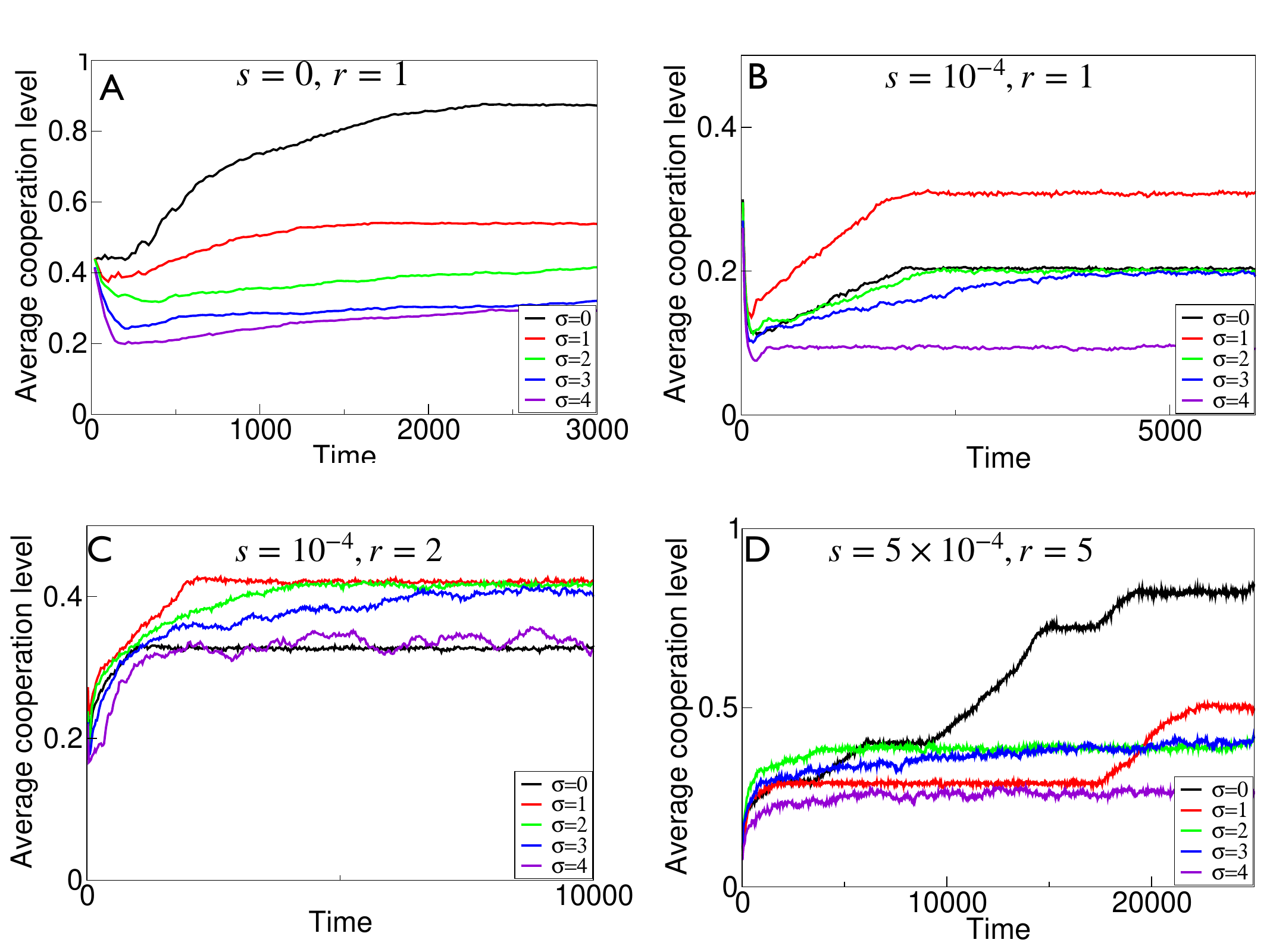} 
   \caption{ The figure illustrates that cooperation can evolve due to spatial structure with the BD update rule. The evolution of the system-average level of cooperation without the within-group games, $s=0$, is presented in panel A.  Panel B shows that a certain level of cooperation is maintained even if within-group games generate selection against cooperation ($s=10^{-4}$). Here the parameter for fissioning randomization was $r=1$, and 
for higher fissioning randomization, generating more variation in the average group trait $\bar x$ among daughter groups, cooperation levels are higher, as shown in panel C ( $s=10^{-4}$, $r=2$) and panel D ($s=5\times 10^{-4}$, $r=5$). {  A completely random choice of the fissioning group, $\s=0$,  results in the highest possible level of cooperation in the case when the within-group games are absent, panel A, or when  the fissioning randomization is strong, panel D. Note that the vertical scales in panels B and C are twice smaller than those in  panels A and D.}} 
   \label{f2}
\end{figure}

\begin{figure}
\centering
  \includegraphics[width=.95\linewidth]{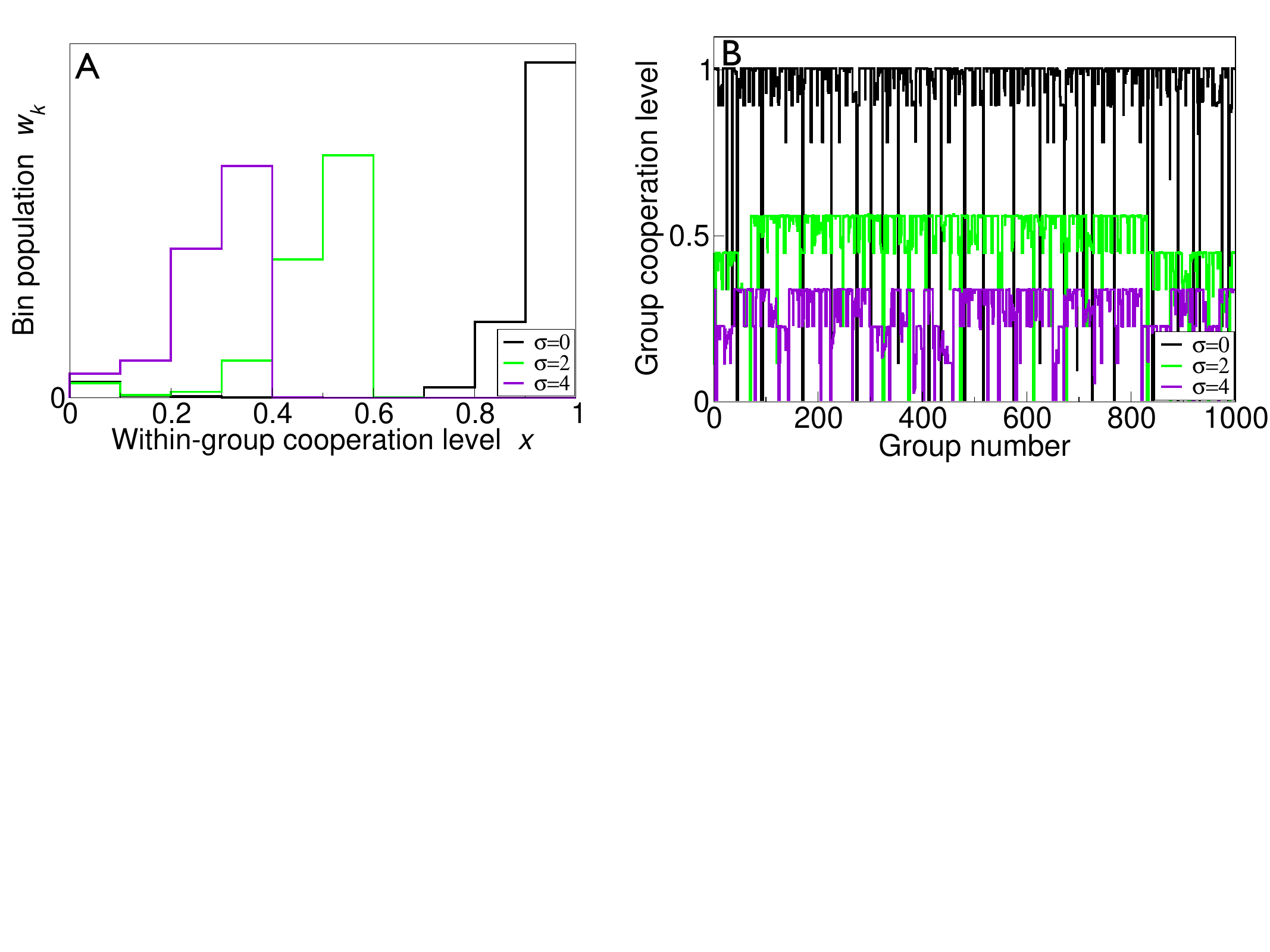} 
   \caption{ Example of the spatial distribution of group cooperation levels, panel A, and of the steady state system-averaged distribution of within-group cooperation levels (bin populations), panel B. BD update, $s=5\times 10^{-4}$ and $r=5$).
} 
\label{f3}
\end{figure}

Consider now the general BD update, where the globally better group, determined probabilistically according to (\ref{softmaxb}), fissions and replaces its worst performing neighbour. In Fig. \ref{f2} we show that cooperation can evolve under this update rule. Cooperation persists not only when the within-group games are turned off ($s=0$, left panel), but also in the presence of within-group selection against cooperation ($s>0$). Note also that without within-group games ($s=0$) the cooperation level is highest when global selection of the fissioning group is neutral, i.e., the choice of fissioning group is random ($\s=0$). However,  to counteract the detrimental effect of within-group games,  ($s>0$),  it is more beneficial to fission better-performing groups, $\s>0$, albeit not the best one: the highest level of cooperation in this case is achieved when the parameter of the softmax function $\s \approx 1$ (panels B,C of Fig. \ref{f2}).
Cooperation persists for a range of within-group selection strength $s$  and fission randomization parameter $r$ as shown in Fig. \ref{f2}. There appears to be a pattern that relates the group selection parameter $\s$ at which cooperation reaches its maximum value to $s$ and $r$.  When the within-group games are relatively strong (larger $s$) and the fission randomization is weak (smaller $r$), cooperation is higher  for non-random selection of the fissioning group, $\s_{max}\in [1,2]$ (Fig. \ref{f2}, panels B, C). On the other hand, when the within-group selection is weaker (smaller $s$) {  or} the fission mixing is strong  (larger $r$), cooperation evolves to higher  levels when selection of the fissioning group is neutral, $\s_{max}=0$ (Fig. \ref{f2}, panels A, D).
Examples of  the system-average distribution of within-group bin populations and the typical equilibrium spatial distributions of group cooperation levels ($\{\bar x\}_i$ vs $i$)  are shown in Fig. \ref{f3}. {  It shows a certain degree of heterogeneity both in within-group and between-group distribution of cooperation level. It makes the corresponding average cooperation levels smaller than the maximum ones achieved by certain individuals or groups.}  

\begin{figure}
\centering
   \includegraphics[width=.95\linewidth]{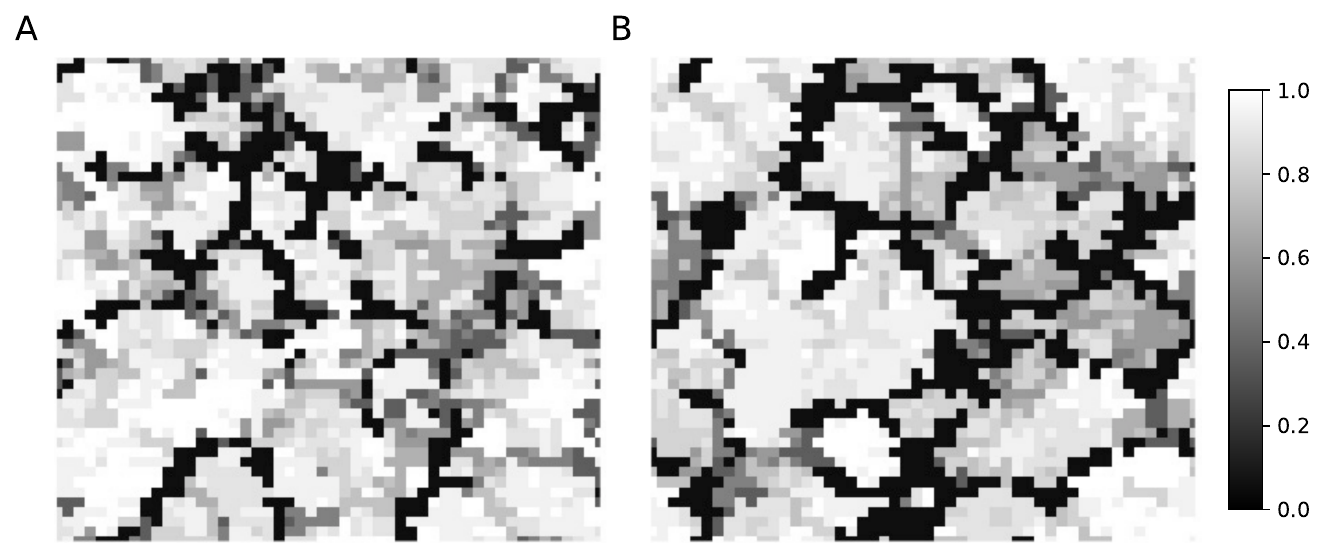} 
  \caption{ The level of cooperation in groups on a 2-dimensional square lattice. Lighter shades of grey correspond to higher cooperation levels. The game parameters are: BD update, $B=4.5$, $C=1$, $r=5$, $\s=1$ in both panels, and $s=2\times 10^{-3}$ and $s=4\times 10^{-3}$ in the left and right panels. The average levels of cooperation at steady state are 0.59 and 0.51 for the left and right panels.}  \label{f6}
\end{figure}

We performed the main part of our analysis with groups embedded in one-dimensional space, but the conclusions qualitatively hold for higher spatial dimensions. As an example, in Fig. \ref{f6} we show the steady state distribution of group cooperation levels on a $50\times50$ square lattice with periodic boundary conditions. Games within groups are the same as in the one-dimensional case, while in space each group plays games with its four nearest neighbours and is updated according to the BD protocol. 

\subsection {DB update}

\begin{figure}
\centering
  \includegraphics[width=.95\linewidth]{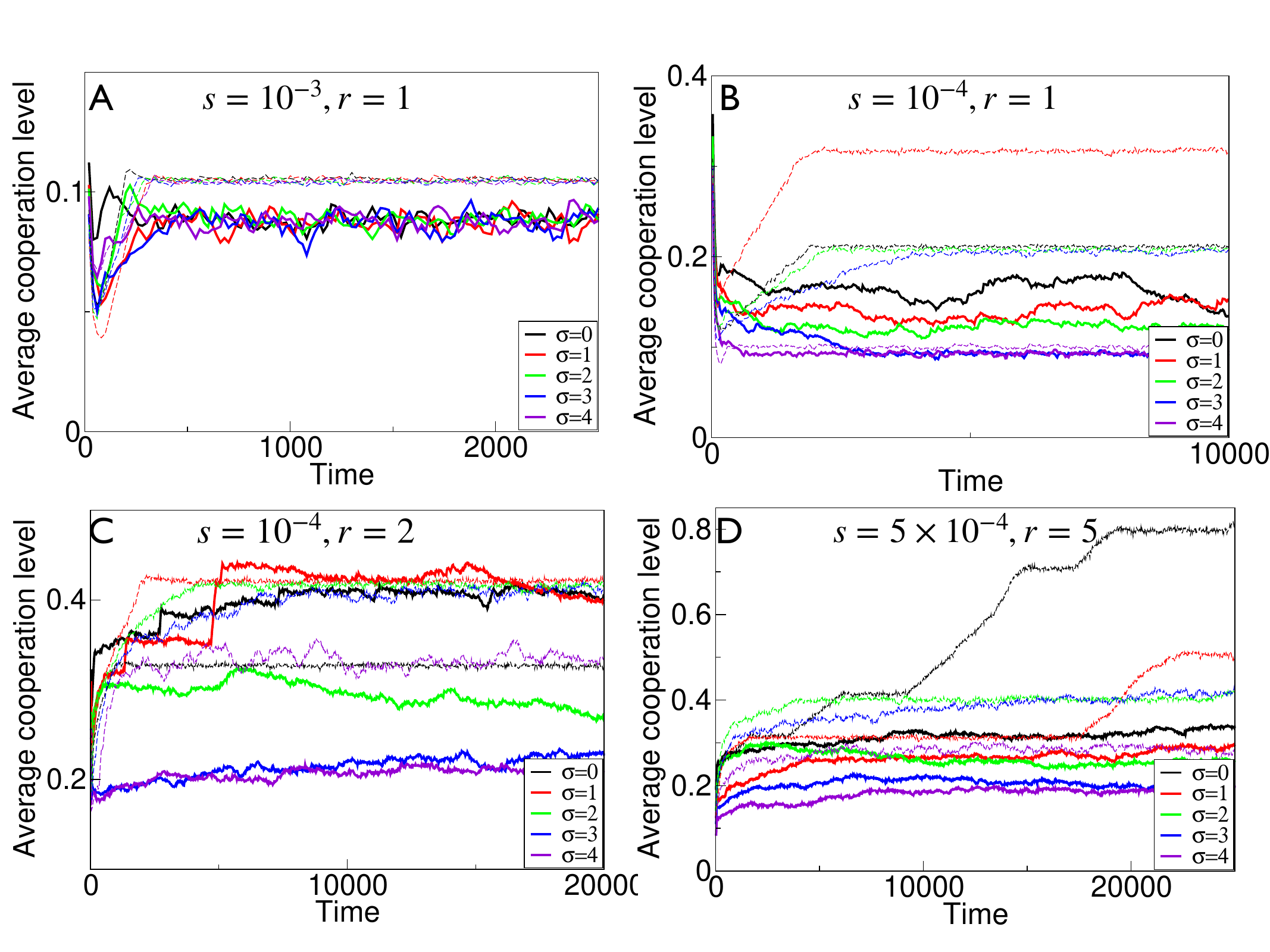}  
  \caption{  The average level of cooperation for the DB update rule: A) $s=10^{-3}$, $r=1$, B) $s=10^{-4}$, $r=1$, C) $s=10^{-4}$, $r=2$, D) $s=5\times10^{-4}$, $r=5$. Thin dashed lines in each panel show the cooperation level for the same parameters under the BD update rule.} 
   \label{f4}
\end{figure}

\begin{figure}
\centering
   \includegraphics[width=.495\linewidth]{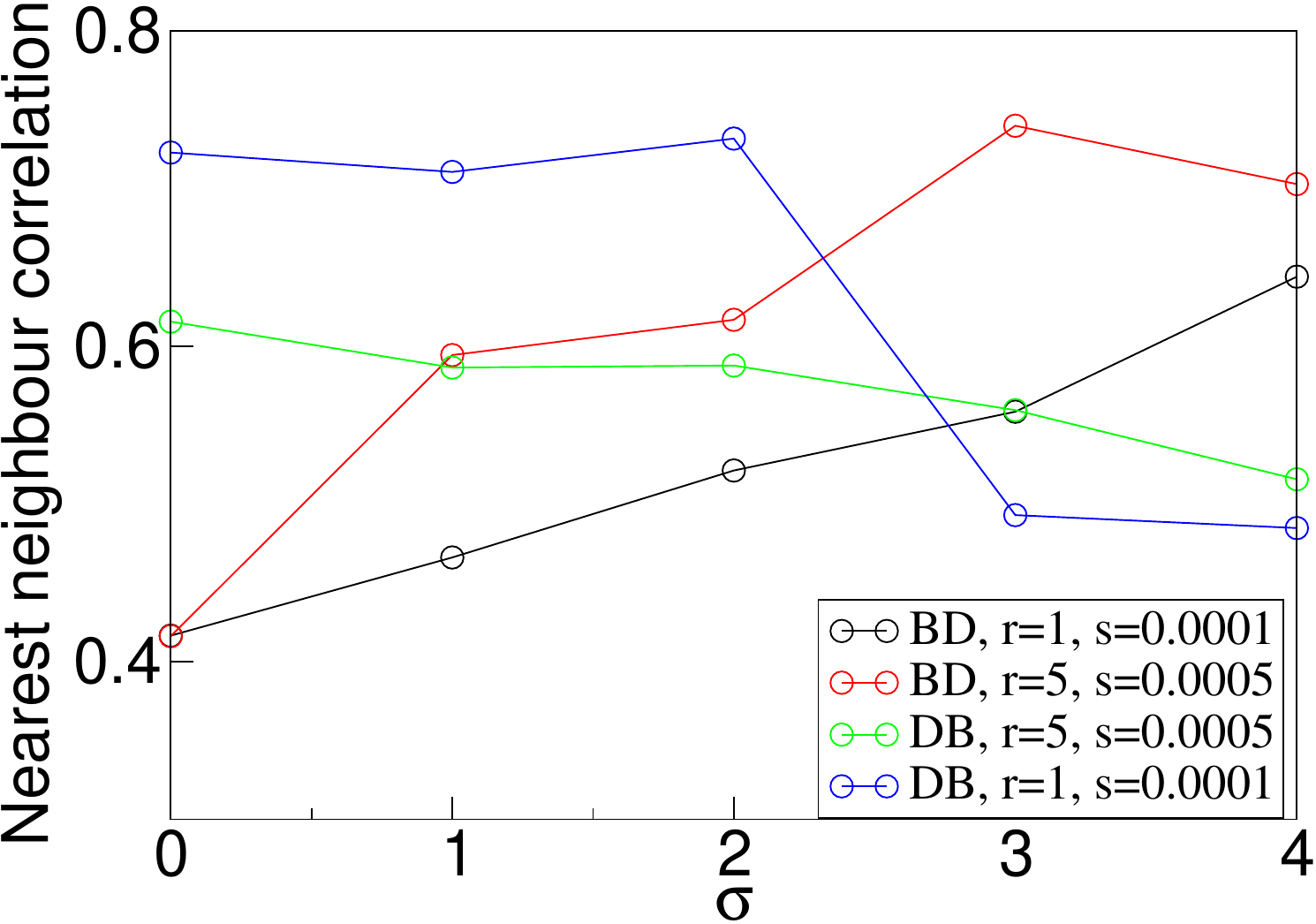} \hfill
   \caption{ Dependence of the nearest neighbour correlation coefficient, $\frac{\sum_i^G \{\bar x\}_i \{\bar x\}_{i+1}}{G Var(\bar x)}$, on the softmax parameter $\s$ for BD and DB update rules.
  } 
   \label{f5}
\end{figure}

Under the Death-Birth update rule, the globally poorly performing group, chosen probabilistically according to (\ref{softmaxw}), is removed and replaced with a fission product of its best-performing neighbour. A random choice of the group to be removed, $\s=0$, represents neutral global selection and corresponds to the traditional DB update in spatial games. We find that with DB updating, cooperation can be maintained for an interval of within-group selection strengths $s$ and fissioning randomization parameters $r$, Fig. \ref{f4}.
While these results qualitatively resemble the conclusions for the BD update and even the onset of maintained non-zero cooperation occurs for approximately the same $r$ and $s$ (panel A), there are several notable differences. First, the average level of cooperation under DB update is somewhat lower than that for the BD update, which is also shown for comparison in all panels in Fig. \ref{f4}  by thin dashed lines.   
Second, under the DB update rule, the highest level of cooperation is achieved when global selection is neutral (for $\s=0$, i.e, when the choice of the group to be removed is random that could be denoted as dB according to \cite{traulsen2023future}), solid black line in all panels in Fig. \ref{f4}. 
It is interesting to see how the difference between the  BD and DB updates is reflected in the correlations between the cooperation levels of neighbouring groups, Fig. \ref{f5}. Maintenance of cooperation invariably implies positive correlations in both cases due to the local character of replacement of the worst group by a fission product of its better-performing neighbour. Yet under the DB update, the strength of global selection $\s$ does not affect or even decreases the nearest neighbour correlation, while under the BD update the correlation increases with $\s$. 

\subsection{Discussion}

Multi-level selection, game theory and spatial structure are all important topics in evolutionary theory. We have long argued that multi-level selection models should be built by incorporating two different birth-death processes, one at the level of individuals, and one at the level of groups, at which birth corresponds to fission and death to extinction \cite{simon2010dynamical,simon2013towards}. Such an approach is very flexible in allowing selection to act on various components of the two-level process. Here we have used this flexibility to study situations in which games are played both between individuals within groups, and between groups in spatially structured populations of groups. We used this setup to study the conditions under which spatial structure can rescue cooperation by assuming that the games played both at the individual and the group level are continuous Prisoner’s Dilemma games. With well-mixed interactions, the continuous Prisoner’s Dilemma always leads to the loss of cooperation, because the only way a player can increase its own payoff is by lowering its cost through decreasing its level of cooperation. Thus, cooperation a priori faces a double whammy in our multi-level selection models: without spatial structure, cooperation is selected against both at the individual and at the group level.

Following classical thinking, we investigated whether spatial structure can reverse this double whammy and enable the maintenance of cooperation. While the within-group games are well-mixed, we assumed that the between-group games are played against nearest neighbours, mostly in a spatially 1-dimensional population of groups. It is known from spatial game theory that the details of the spatial update rules are important, and in agreement with classical results, we find that cooperation cannot be maintained if the globally best-performing group is chosen to fission with one of the fissioning products replacing a randomly chosen neighbour (and the other replacing the parent group). This result is know as the cancellation effect in classical 1-level spatial game theory \cite{taylor1992altruism, wilson1992can}. However, in virtually every other update scenario that we tested, some level of cooperation can be maintained in the spatially structured 2-level model. Thus, in both the Birth-Death (BD) and the Death-Birth (DB) scenarios, in each of which selection due to games between groups acts both on fissioning and extinction rates to various degrees, cooperation can persist for a range of relative strengths of between-group and within-group games. The parameter controlling  the randomization during group fissioning ($r$) also affects the dynamics, but even large variation generation cannot overcome the cancellation effect in the BD scenario with random local extinction. 

Generalizing the cancellation effect in single-level spatial games, we find that certain selection scenarios are particularly favourable for the maintenance of cooperation in our spatial 2-level models. Specifically, both in the BD and the DB scenarios, in which a global first step is followed by a local second step, selection in the global step is not necessary for the maintenance of cooperation.
In our model, the strength of the global selection is controlled by the parameters $\s$, which determines how strict the choice of the system-wide best-performing or worst-performing group is. When $\s=0$, global selection is neutral, while for large $\s$, the absolute best- or worst-performing group is chosen. We observed that for the BD selection, the highest cooperation level evolves for $\s=0$ in the case of weak within-group selection (small $s$) or strong fissioning randomization (large $r$), while in the DB scenarios, 
globally neutral selection($\s=0$) is always the most beneficial for cooperation. On the other hand, local selection 
is always essential for the evolution of cooperation. As mentioned, BD updating with neutral local selection (corresponding to the classical Bd scheme), prevents the evolution of cooperation in the spatial 1-level Prisoner’s Dilemma. Similarly, neutral local selection also prevents cooperation under Db updating. Therefore, the common principle emerges that local selection is crucial for the maintenance of cooperation, while global selection is optional. {  The effect of the within-group games strength $s$ and randomization parameter $r$ on the outcome of two-level games are qualitatively similar to those reported for non-spatial scenarios in \cite{simon2024evolutionary,simon2024fission}. Stronger within-group selection predictably makes those games more influential and reduces the overall cooperation, while the stronger randomization increases the between-group variance and consequently strengthens the between-group games, increasing the cooperation.}

\begin{sloppypar}
There appear to be few examples of spatially-structured multi-level models involving the Prisoner’s Dilemma game in the literature, e.g. \cite{akdeniz2020cancellation}. Indeed, \cite{akdeniz2020cancellation} have observed the cancellation effect in their model. However, these authors employed a much simpler selection scenarios at the group level, assuming directional selection on cooperation levels, which by itself of course enhances rather than diminishes cooperation. Our approach involving evolutionary games at the group level can easily be generalized to situations where games other than the (continuous) Prisoner’s Dilemma game are played between groups. One worthwhile case would be the Continuous Snowdrift game \cite{hauert2004spatial, doebeli2005models, simon2024evolutionary}. It is known that spatial structure actually 	inhibits cooperation in the 1-level spatial Snowdrift game \cite{hauert2004spatial}, and it would therefore be interesting to investigate spatially structured 2-level Snowdrift games. The problem here is that even the well-mixed 1-level Snowdrift game is much more complicated than the well-mixed 1-level Prisoner’s Dilemma, exhibiting much richer evolutionary dynamics \cite{doebeli2005models}. Not surprisingly then, the 2-level well-mixed Snowdrift game is also rather more complicated than the 2-level well-mixed Prisoner’s Dilemma \cite{simon2024evolutionary}. Therefore, 2-level spatial Snowdrift games will require a careful analysis, which we leave for future work.
\end{sloppypar}

In summary, we have provided yet another set of examples where group-level processes can rescue cooperation that is doomed in the corresponding individual-level process. What is new here is that cooperation is in fact also doomed in the group-level process if selection at the group level occurs in well-mixed populations of groups. Only spatial structure in the population of groups allows the group-level process to rescue cooperation. It has recently been argued that there is ample empirical evidence for the importance of multi-level selection \cite{marin2026abundant}, and we believe that it is of general importance to develop and expand multi-level selection theory.


\section*{Funding}
YI acknowledges support from DICYT USACH project no. 42531ISSA REG.  
\section* {Conflict of interest}
The authors declare no conflict of interest.

\bibliography{space.bib}

@article{nowaksigmund2000grids,
  author    = {Nowak, M. A. and Sigmund, K. },
  title     = {Games on Grids},
  booktitle = {The Geometry of Ecological Interactions:Simplifying Spatial Complexity},
  publisher = {Cambridge University Press},
  year      = {2000},
  editor    = {Dieckmann U. and Law, R. and Metz, J.A.J.},
  pages     = {135--150},
}

@article{kerrgodfrey2002mls,
  title={Individualist and Multi-level Perspectives on Selection
in Structured Populations},
  author={Kerr, B. and Godfrey-Smith, P.},
  journal={Biology and Philosophy 17: 477–517, 2002},
  volume={17},
  pages={477--517},
  year={2002},
}

@article{champagnat2006unifying,
  title={Unifying evolutionary dynamics: from individual stochastic processes to macroscopic models},
  author={Champagnat, Nicolas and Ferri{\`e}re, R{\'e}gis and M{\'e}l{\'e}ard, Sylvie},
  journal={Theoretical population biology},
  volume={69},
  number={3},
  pages={297--321},
  year={2006},
  publisher={Elsevier}
}

@book{hofbauersigmund2012games,
  title={Evolutionary games and population dynamics},
  author={Hofbauer, J. and Sigmund, K.},
  year={2012},
  publisher={Cambridge University Press}
}

@article{luo2014unifying,
  title={A unifying framework reveals key properties of multilevel selection},
  author={Luo, Shishi},
  journal={Journal of Theoretical Biology},
  volume={341},
  pages={41--52},
  year={2014},
  publisher={Elsevier}
}

@article{simon2010dynamical,
  title={A dynamical model of two-level selection},
  author={Simon, Burton},
  journal={Evolutionary Ecology Research},
  volume={12},
  number={5},
  pages={555--588},
  year={2010},
  publisher={Evolutionary Ecology, Ltd.}
}

@article{simon2013towards,
  title={Towards a general theory of group selection},
  author={Simon, Burton and Fletcher, Jeffrey A and Doebeli, Michael},
  journal={Evolution},
  volume={67},
  number={6},
  pages={1561--1572},
  year={2013},
  publisher={Blackwell Publishing Inc Malden, USA}
}

@article{simon2012numerical,
  title={Numerical solutions and animations of group selection dynamics},
  author={Simon, Burton and Nielsen, Aaron},
  journal={Evolutionary Ecology Research},
  volume={14},
  number={6},
  pages={757--768},
  year={2012},
  publisher={Evolutionary Ecology, Ltd.}
}

@article{simon2016group,
  title={Group-level events are catalysts in the evolution of cooperation},
   author={Simon, Burton and Pilosov, Michael},
  journal={Journal of Theoretical Biology},
  volume={410},
  pages={125--136},
  year={2016},
  publisher={Elsevier}
}

@article{traulsen2006evolution,
  title={Evolution of cooperation by multilevel selection},
  author={Traulsen, Arne and Nowak, Martin A},
  journal={Proceedings of the National Academy of Sciences},
  volume={103},
  number={29},
  pages={10952--10955},
  year={2006},
  publisher={National Acad Sciences}
}

@article{nowak1992evolutionary,
  title={Evolutionary games and spatial chaos},
  author={Nowak, Martin A and May, Robert M},
  journal={Nature},
  volume={359},
  number={6398},
  pages={826--829},
  year={1992},
  publisher={Nature Publishing Group UK London}
}

@article{cooney2019replicator,
  title={The replicator dynamics for multilevel selection in evolutionary games},
  author={Cooney, Daniel B},
  journal={Journal of mathematical biology},
  volume={79},
  number={1},
  pages={101--154},
  year={2019},
  publisher={Springer}
}

@article{cooney2022long,
  title={Long-time behavior of a PDE replicator equation for multilevel selection in group-structured populations},
  author={Cooney, Daniel B and Mori, Yoichiro},
  journal={Journal of Mathematical Biology},
  volume={85},
  number={2},
  pages={12},
  year={2022},
  publisher={Springer}
}

@article{taylor1992altruism,
  title={Altruism in viscous populations—an inclusive fitness model},
  author={Taylor, Peter D},
  journal={Evolutionary ecology},
  volume={6},
  number={4},
  pages={352--356},
  year={1992},
  publisher={Springer}
}

@article{akdeniz2020cancellation,
  title={The cancellation effect at the group level},
  author={Akdeniz, Asl{\i}han and van Veelen, Matthijs},
  journal={Evolution},
  volume={74},
  number={7},
  pages={1246--1254},
  year={2020},
  publisher={Blackwell Publishing Inc Malden, USA}
}

@article{ohtsuki2006simple,
  title={A simple rule for the evolution of cooperation on graphs and social networks},
  author={Ohtsuki, Hisashi and Hauert, Christoph and Lieberman, Erez and Nowak, Martin A},
  journal={Nature},
  volume={441},
  number={7092},
  pages={502--505},
  year={2006},
  publisher={Nature Publishing Group UK London}
}

@article{taylor2007evolution,
  title={Evolution of cooperation in a finite homogeneous graph},
  author={Taylor, Peter D and Day, Troy and Wild, Geoff},
  journal={Nature},
  volume={447},
  number={7143},
  pages={469--472},
  year={2007},
  publisher={Nature Publishing Group UK London}
}

@article{hauert2021spatial,
  title={Spatial social dilemmas promote diversity},
  author={Hauert, Christoph and Doebeli, Michael},
  journal={Proceedings of the National Academy of Sciences},
  volume={118},
  number={42},
  pages={e2105252118},
  year={2021},
  publisher={National Academy of Sciences}
}

@article{zukewich2013consolidating,
  title={Consolidating birth-death and death-birth processes in structured populations},
  author={Zukewich, Joshua and Kurella, Venu and Doebeli, Michael and Hauert, Christoph},
  journal={PLoS One},
  volume={8},
  number={1},
  pages={e54639},
  year={2013},
  publisher={Public Library of Science San Francisco, USA}
}

@article{simon2024fission,
  title={Fission as a source of variation for group selection},
  author={Simon, Burton and Ispolatov, Yaroslav and Doebeli, Michael},
  journal={Evolution},
  volume={78},
  number={9},
  pages={1583--1593},
  year={2024},
  publisher={Oxford University Press US}
}

@article{simon2024evolutionary,
  title={Evolutionary branching in multi-level selection models},
  author={Simon, Burton and Ispolatov, Yaroslav and Doebeli, Michael},
  journal={Journal of Mathematical Biology},
  volume={89},
  number={5},
  pages={52},
  year={2024},
  publisher={Springer}
}

@article{axelrod1981evolution,
  title={The evolution of cooperation},
  author={Axelrod, Robert and Hamilton, William D},
  journal={Science},
  volume={211},
  number={4489},
  pages={1390--1396},
  year={1981},
  publisher={American Association for the Advancement of Science}
}

@article{sigmund1999evolutionary,
  title={Evolutionary game theory},
  author={Sigmund, Karl and Nowak, Martin A},
  journal={Current Biology},
  volume={9},
  number={14},
  pages={R503--R505},
  year={1999},
  publisher={Elsevier}
}

@article{traulsen2023future,
  title={The future of theoretical evolutionary game theory},
  author={Traulsen, Arne and Glynatsi, Nikoleta E},
  journal={Philosophical Transactions of the Royal Society B: Biological Sciences},
  volume={378},
  number={1876},
  year={2023},
  publisher={The Royal Society}
}

@article{doebeli2005models,
  title={Models of cooperation based on the Prisoner's Dilemma and the Snowdrift game},
  author={Doebeli, Michael and Hauert, Christoph},
  journal={Ecology letters},
  volume={8},
  number={7},
  pages={748--766},
  year={2005},
  publisher={Wiley Online Library}
}

@article{hauert2004spatial,
  title={Spatial structure often inhibits the evolution of cooperation in the snowdrift game},
  author={Hauert, Christoph and Doebeli, Michael},
  journal={Nature},
  volume={428},
  number={6983},
  pages={643--646},
  year={2004},
  publisher={Nature Publishing Group UK London}
}

@article{marin2026abundant,
  title={Abundant empirical evidence of multilevel selection revealed by a bibliometric review},
  author={Mar{\'\i}n, C{\'e}sar and Clark, Anne B and Philson, Conner S and Eldakar, Omar Tonsi and Wade, Michael J},
  journal={Frontiers in Ecology and Evolution},
  volume={14},
  pages={1752597},
  year={2026},
  publisher={Frontiers Media SA}
}

@article{cooney2023evolutionary,
  title={Evolutionary dynamics within and among competing groups},
  author={Cooney, Daniel B and Levin, Simon A and Mori, Yoichiro and Plotkin, Joshua B},
  journal={Proceedings of the National Academy of Sciences},
  volume={120},
  number={20},
  pages={e2216186120},
  year={2023},
  publisher={National Academy of Sciences}
}

@article{fog2026simulation,
  title={Simulation of group selection models},
  author={Fog, Agner},
  journal={Bulletin of Mathematical Biology},
  volume={88},
  number={2},
  pages={27},
  year={2026},
  publisher={Springer}
}

@article{wilson1992can,
  title={Can altruism evolve in purely viscous populations?},
  author={Wilson, David Sloan and Pollock, Gregory B and Dugatkin, Lee A},
  journal={Evolutionary ecology},
  volume={6},
  number={4},
  pages={331--341},
  year={1992},
  publisher={Springer}
}
\bibliographystyle{apalike}     

\section*{Appendix}
\renewcommand{\theequation}{A.\arabic{equation}}
\setcounter{equation}{0}
\subsection*{Decay of cooperation due to the within-group dynamics}
Here we derive a deterministic approximation under weak selection for the evolution of the average level of cooperation $\bar x$ given by the within-group PDE (\ref{integdiffeqn}),
\begin{align}
\label{def}
\bar x = \frac {\int n(x) x dx}{\int n(x) dx}=\eta^{-1}\int n(x) x dx
 \end {align}  
We assume that the total within-group population size $\eta$ has relaxed to its equilibrium value $\b/\d$ and no longer depends on time.
 From (\ref{integdiffeqn}) it follows that
\begin{align}
\label{ode1}
\eta \frac{d \bar x}{dt} = \b\eta \frac {\int n(x) e^{sP(x)}x dx}{\int n(x) e^{sP(x)} dx} - \d \bar x \eta^2
\end {align}  
Assuming weak selection, $e^{sP(x)}$ can be approximated to first order as
\begin{align}
\label{weak}
  e^{sP(x)}\approx 1+ sP(x)=1- sCx + sB\bar x.
\end {align}  
Substitution of  (\ref{weak}) into (\ref{ode1}) and subsequent integration results in 
\begin{align}
\label{ode2}
 \frac{d \bar x}{dt} = \beta \frac{\bar x - sC\langle x^2 \rangle + sB {\bar x}^2}{1-sC\bar x + sB \bar x} - \d \eta \bar x,
\end {align}
where $\langle x^2 \rangle $ is the second moment of $x$.
Taking into account the equilibrium population size 
$\eta=\b/\d$,
Eq. (\ref{ode2}) yields
\begin{align}
\label{ode3}
 \frac{d \bar x}{dt} = -\beta s \frac{C {\sigma^2_x}}{1+s\bar x(B-C)}.
  \end {align}
 Here ${\sigma^2_x}$ is the variance of the cooperation level $x$. 
This expression is negative for social dilemmas ($B>C$). It confirms that when considering groups in isolation, cooperation within groups decays when individuals are playing the continuous Prisoner’s Dilemma game. It also shows that the selection strength is proportional to the product $\beta s$.

\end{document}